# Equation of state for hard sphere fluids across the stable and metastable regimes

Can Cui[a] & Jianxiang Tian[ab*]

[a] Department of Physics, Qufu Normal University, Qufu 273165, P. R. China

[b] Department of Physics, Dalian University of Technology, Dalian 116024, P. R. China

*jxtian@qfnu.edu.cn; jxtian@dlut.edu.cn

**Abstract**

We review selected equations of state (EOSs) for three-dimensional hard-sphere fluids and revise a closed-virial construction to describe stable and metastable disordered states. The equation preserves the prescribed coefficients $B_2 - B_{11}$, uses a linear continuation constrained by estimates of $B_{12}$–$B_{16}$, and imposes a simple pole at the representative maximally random jammed packing fraction $\eta_p = 0.637$. The pole is a modeling assumption for a disordered continuation and is not identified with an exact singularity of the equilibrium virial series. The remaining parameters are fitted to the high-density subset of the simulation data of Pieprzyk et al. [Phys. Chem. Chem. Phys. 21, 6886 (2019)]. The overall average absolute relative deviation is 0.02236% over their 100 data points. With unchanged parameters, the deviation is 0.005874% over 50 points from Davidchack and Laird [J. Chem. Phys. 157, 074701 (2022)]. Agreement with the high-order coefficient estimates is assessed as a construction consistency check. Comparisons distinguish the information used to determine each EOS from its subsequent performance. The resulting equation combines prescribed virial inputs, a representative disordered endpoint, and small pressure deviations over the tested range, while retaining a phenomenological high-order continuation.

## 1. Introduction

Hard spheres (HSs) provide a simple yet fundamental model for studying the physical properties of three-dimensional gases, liquids, and solids[1, 2]. The interaction between two hard spheres is zero when their center-to-center distance exceeds the sum of their radii, and infinite when this distance is smaller than the sum of their radii, corresponding to particle overlap. Owing to its conceptual simplicity and physical relevance, the HS model is one of the most widely used reference models for describing fluid behavior[1, 2], particularly in statistical associating fluid theory[3] and perturbation theory[4]. It has also provided an important foundation for advances in the study of real liquids, amorphous liquids and solids, liquid crystals, granular matter, and model colloidal systems[1, 5-7].

An accurate hard-sphere EOS provides pressure and, through thermodynamic integration or differentiation, other fluid properties such as excess free energy, chemical potential, and compressibility.[8–12] The compressibility factor is written as

$$Z = \frac{P}{k_B T \rho} = \frac{P^*}{\eta} = 1 + B_2\eta + B_3\eta^2 + \cdots \quad (1)$$

where $Z$ is called the compressibility factor, $P$ is the pressure, $k_B$ is the Boltzmann constant, $T$ is the temperature, $\rho$ is the number density defined as $\rho = N/V$ with $N$ being the number of particles and $V$ being the volume, $P^*$ is the reduced pressure defined as $Pv/(k_B T)$ with $v$ being the volume of a hard sphere, $B_j$ is the $j$th volume-reduced virial coefficient, and $\eta = y = \pi\rho\sigma^3/6$ is the packing fraction, and $\sigma$ is the hard-sphere diameter and is taken as the unit of length. Both $Z$ and $\eta$ are dimensionless quantities. $Z = 1$ corresponds to the EOS of the ideal gas, *i.e.*, $PV = Nk_B T$.

Accurate virial coefficients and simulation pressures provide complementary constraints for an EOS. We consider the additional modeling objective of a finite-density divergence for a metastable disordered continuation. These objectives need not select the EOS with the smallest pressure residuals. Several empirical equations are exceptionally accurate over the available data but diverge at $\eta = 1$, whereas imposing a representative disordered endpoint can reduce fitting flexibility. We therefore assess the balance among reproduction of prescribed low-order coefficients, agreement with simulation pressures, and a chosen high-density continuation. The present work updates an existing closed-virial framework rather than deriving a unique EOS from microscopic theory. For the broader historical survey, we refer to Mulero et al.[7] Our selected comparisons emphasize the forms and constraints relevant to this construction and are not an exhaustive catalogue. Section 2 describes the reference information, Sec. 3 presents the benchmark equations, Sec. 4 develops the revised EOS, and Secs. 5 and 6 give the assessment and conclusions.

## 2. Numerical values of virial coefficients and the compressibility factor

Virial coefficients are important because they are directly related to the intermolecular potential energy function of the molecules concerned[12]. For real gases and liquids, the virial coefficients are generally temperature dependent[12, 13]. In contrast, the virial coefficients of hard-sphere fluids are independent of temperature, owing to the purely excluded-volume nature of the hard-sphere interaction. The second through fourth virial coefficients, $B_2$-$B_4$, were derived exactly by Boltzmann and subsequent studies[1, 14-16]. For higher order virial coefficients, exact analytical values are generally unavailable because of the complexity of the associated multidimensional integrals, and numerical methods are therefore required.

Numerical calculations of hard-sphere virial coefficients have progressed from early Monte Carlo estimates to high-order cluster-integral methods.[17–26] Table 1 compiles exact analytical values, numerical estimates, and extrapolated estimates, which must be distinguished. We retain the input values of $B_2$–$B_{11}$ used in Refs. 8 and 9 and the estimates of $B_{12}$–$B_{16}$ from Ref. 22. The numerical inputs above fourth order have uncertainties; reproducing them exactly does not make them exact physical constants. In particular, the $B_{12}$ values of 131 ± 23 and 133 ± 23 in Table 1 overlap within their reported uncertainties. Their small difference is not evidence of statistical inconsistency. The construction uses an explicit polynomial through $B_{11}$ and treats the less certain higher-order information separately.

To assess the accuracy of a HS EOS, reliable numerical data for the compressibility factor $Z$ as a function of number density $\rho$, or equivalently packing fraction $\eta$, are required. Such data are typically obtained from Monte Carlo (MC) and molecular dynamics (MD) simulations. Simulation results published before 2000—including those of Alder and Wainwright[27] in 1960, Alder, Hoover and Young[28] in 1968, Hoover and Ree[29] in 1968, Barker and Henderson[30] in 1971, Adams[31] in 1974, Woodcock[32] in 1976, Labik and Malijevsky[33] in 1981, and Erpenbeck and Wood[34] in 1984–were collected and analyzed by Wu and Sadus[35]. More recent data published after 2000, including those of Kolafa, Labik and Malijevsky[25] in 2004, Wu and Sadus[35] in 2005, Bannerman, Lue and Woodcock[15] in 2010, Irrgang *et al*[36] in 2017, and Pieprzyk *et al*[37] in 2019, were subsequently collected and analyzed by Tian *et al*[8]. In 2022, Davidchack and Laird reported a new set of highly precise simulation data[11].

The principal dataset comprises 100 simulation points from Pieprzyk et al.,[37] spanning $0.050 \le \rho^* \le 1.020$, where $\rho^* = \rho\sigma^3$ and $\eta = \pi\rho^*/6$. The corresponding packing fractions extend from 0.02617994 to 0.53407075. We use the 41 points with $\rho^* \ge 0.940$ to determine the remaining EOS parameters. The 59 lower-density points provide an additional check within this dataset. A separate evaluation uses 50 Davidchack–Laird points[11] at $\eta = 0.01, 0.02, \dots, 0.50$, without refitting the present EOS. This second dataset tests transfer to another set of pressures but does not extend the maximum density tested. The empirical DL equation was fitted using data from both sources, and its residuals must be interpreted accordingly.

## 3. Benchmark EOSs of hard sphere fluids and design considerations

The *i*th order virial EOS reads[36, 38]

$$Z_{VEOSi} = 1 + \sum_{l=2}^{i} B_l \eta^{l-1} \quad (2)$$

This equation is a truncated part of the infinite series Eq. (1) and was used to check the effective order of virial coefficients for describing the properties of both the HS fluid systems and real fluid systems[1, 36, 38]. We emphasize again that the virial coefficients are temperature dependent for real fluid systems. As a result, the researches on the accurate correlations of virial coefficients versus the temperature, for instance $B_2=B_2(T)$ and $B_3=B_3(T)$, are very important for effectively constructing the *i*th order virial EOS of real fluid systems. See Refs. ([13, 39, 40]) for details and references therein.

The *j*th order exponential approximant (EA*j*) EOS for the HS fluids proposed by Barlow *et al*. reads[41]

$$Z_{EAj} = \exp(N_2\eta + N_3\eta^2 + \cdots + N_j\eta^{j-1}) \quad (3)$$

Originally, Barlow *et al* proposed a generalized Pade approximant for repulsive spheres of arbitrary softness[41]. It is found that the effectiveness of the approximant is well enhanced over the conventional Pade approximants through enforcing the same high-density asymptotic behavior as the model fluid being described[41]. Eq. (3) is the HS limit of their approximants. The coefficients $N_j$ in Eq. (3) can be determined by matching the Taylor expansion coefficients of $Z_{EAj}$ to the known virial coefficients of HS fluids and numerically solving the corresponding equation systems by Maple or Matlab software. In this paper, both $Z_{EA11}$ and $Z_{EA12}$ are included for analysis:

$$Z_{\mathrm{EA11}} = \exp\begin{pmatrix} 4\eta + 2\eta^2 - 0.3018982838\eta^3 + 0.7653034686\eta^4 \\ +1.906332297\eta^5 - 0.7019560856\eta^6 - 0.2961142343\eta^7 \\ +3.023625320\eta^8 - 0.8423249606\eta^9 - 3.639942812\eta^{10} \end{pmatrix} \quad (4)$$

$$Z_{\mathrm{EA12}} = \exp\begin{pmatrix} 4\eta + 2\eta^2 - 0.3018982838\eta^3 + 0.7653034686\eta^4 \\ +1.906332297\eta^5 - 0.7019560856\eta^6 - 0.2961142343\eta^7 \\ +3.023625320\eta^8 - 0.8423249606\eta^9 - 3.639942812\eta^{10} \\ -11.34722973\eta^{11} \end{pmatrix} \quad (5)$$

Eqs. (4-5) were reported by Tian *et al* through using their best estimate values of virial coefficients[8].

The truncated polynomial, Eq. (2), and the exponential of a finite polynomial, Eq. (3), have no finite pole. A geometric high-order coefficient continuation associated with a prescribed positive pole has the limiting ratio discussed in Ref. 42,

$$\lim_{n\to\infty}(B_n/B_{n+1}) = \eta_p < 1 \quad (6)$$

Equation (6) is used here to characterize the asymptotic behavior imposed on the model. It does not establish the singularity structure of the exact hard-sphere virial series. A positive limiting ratio does not determine the sign of every coefficient, and it does not require monotonic growth at a specified finite order. Consequently, a negative coefficient or a finite-order decrease is not, by itself, evidence of a violation of Eq. (6). The Carnahan–Starling equation provides a familiar example with a different pole position, $\eta_p = 1$.[43]

Following, we show some published HS EOSs with non-zero poles. The famous CS EOS reads[43]

$$Z_{CS} = \frac{1+\eta+\eta^2-\eta^3}{(1-\eta)^3} \quad (7)$$

The virial coefficients from CS EOS are all integers and its packing fraction pole reads $\eta_p = 1$.

Kolafa's modification of the Carnahan–Starling equation (CSK) was reported in Eq. (4.46) of Boublík and Nezbeda[38] and is also given as Eq. (2.6) by Heyes and Santos:[44]

$$Z_{CSK} = \frac{1+\eta+\eta^2-(2/3)\eta^3(1+\eta)}{(1-\eta)^3} \quad (7a)$$

This parameter-free modification retains a third-order pole at $\eta$=1. Its virial coefficients are $B_2 = 4$ and $B_n = (5/6)(n^2 + 3n - 6)$ for $n \geq 3$.

Based on the Percus-Yevick integral equation, Sun *et al*[45] proposed a universal cubic (UC) EOS as

$$Z_{UC} = 1 + \frac{4\eta}{1-1.126\eta} + \frac{5.696\eta^2}{(1-1.126\eta)^2} \tag{8}$$

The authors used the simulation data of $Z$ versus $\eta$ by Bannerman *et al*[15] to test the accuracy of equations considered, and found that this UC EOS gives the third best results and better than the Carnahan-Starling EOS. $\eta_p$ of this UC equation is 0.888.

Through integrating the isothermal compressibility $\chi_T$ starting from the Percus-Yevick closure of the Ornstein-Zernike integral equation, Hansen-Goos[46] derived an equation of state for HS fluids with its final form being

$$Z_H = \frac{a\ln(1-\eta)}{\eta} + \frac{\sum_{i=0}^{7} b_i \eta^i}{(1-\eta)^3} \tag{9}$$

The coefficients are $a = 8$, $b_0 = 9$, $b_1 = -19$, $b_2 = 47/3$, $b_3 = -2.635232$, $b_4 = -1.265575$, $b_5 = 0.041212$, $b_6 = 0.248245$, and $b_7 = -0.096495$. The equation is accurate over the stable-fluid range considered in Ref. 46. Its finite positive singularity is at $\eta = 1$. The apparent division by $\eta$ in the logarithmic term is removable at the origin because $\ln(1-\eta)/\eta \to -1$.

Bonneville[47] proposed a semi empirical HS EOS as

$$Z_B = 1 + 4\eta + 10\eta^2 \left(\frac{1}{1-\frac{\eta}{\eta_0}}\right) \frac{1-1.7343\eta+0.4063\eta^2}{(1-\eta)^2} \tag{10}$$

with $\eta_0 = 0.6374$. Bonneville[47] declared that Eq. (10) is valid in both the stable and metastable phases. Clearly, this EOS has two poles as $\eta_p = 1$ and $\eta_p = 0.6374$.

Based on the asymptotic expansion method[48, 49] (AEM), Tian *et al*[14] proposed the following HS EOS

$$Z_{\text{AEM}} = Z(i,j) = \sum_{k=i}^{j} a_k X^k, \quad j > i, \quad i,j,k \in \mathbb{Z}. \tag{11}$$

In Eq. (11), coefficients $a_k$ are determined by known virial coefficients, $X = 1/(\eta - b)$ with $b$ being the radius of convergence of the virial expansion, *i. e.*, the packing pole $\eta_p$. Let $b$=1 and take the integer values of the first three virial coefficients, Eq. (11) reads[14] the Carnahan-Starling equation[43]. By using the numerical values for the first ten virial coefficients, the authors finally selected $Z$(-5, 2) out of 57 possible equations. $Z$(-5, 2) was shown that it is less accurate than other EOSs (such as Santos *et al*[50], Kolafa *et al*[25]) at the stable densities, and of similar accuracy to most of the others in the metastable densities. Its main advantage is that, by using the first ten virial coefficients, it is capable of reproducing all the 16 virial coefficients (the first ten numerical and the others predicted by Clisby and McCoy[22]) locating in the error regions. The updated version of Eq. (11) by using the best estimate values of the first 12 virial coefficients in **T**able 1 is specified in Refs. 8 and 9, including the published correction

with $a_{-5} = -0.2418392938$ ; $a_{-4} = -1.364701188$ ; $a_{-3} = -3.018299598$ ; $a_{-2} = -2.974598552$ ; $a_{-1} = 0.6225237445$ ; $a_0 = 6.80859329042177$ ; $a_1 = 9.714037872$ ; $a_2 = 5.355509364$ ; $b = 0.9246639546$. $\eta_p$ of this equation is 0.9247.

Padé approximants [4/5], [5/4], [5/5], [5/6], and [6/5] by using the best estimate values of virials by Tian *et al* are as follows[8]

$$P[4/5] = \frac{1.115861884\eta^4 + 2.766955702\eta^3 + 3.656210148\eta^2 + 2.221211866\eta + 1}{-0.5277931041\eta^5 + 1.426418437\eta^4 - 0.8953820783\eta^3 + 0.7713626843\eta^2 - 1.778788134\eta + 1} \quad (12)$$

$$P[5/4] = \frac{2.794177294\eta^5 + 7.551568933\eta^4 + 9.174783436\eta^3 + 7.556546765\eta^2 + 3.304302827\eta + 1}{3.071574989\eta^4 - 3.590355044\eta^3 + 0.3393354571\eta^2 - 0.6956971731\eta + 1} \quad (13)$$

$$P[5/5] = \frac{2.954324373\eta^5 + 7.920428736\eta^4 + 9.542045347\eta^3 + 7.780092893\eta^2 + 3.366379723\eta + 1}{0.03025023650\eta^5 + 3.165866435\eta^4 - 3.744816270\eta^3 + 0.3145740010\eta^2 - 0.6336202770\eta + 1} \quad (14)$$

$$P[5/6] = \frac{-10.47704261\eta^5 - 25.38476610\eta^4 - 34.46689605\eta^3 - 18.95611178\eta^2 - 8.670385018\eta + 1}{6.352921425\eta^6 - 17.13921291\eta^5 + 13.94336986\eta^4 - 13.02952743\eta^3 + 21.72542829\eta^2 - 12.67038502\eta + 1} \quad (15)$$

$$P[6/5] = \frac{-620.4444256\eta^6 - 1673.864416\eta^5 - 2029.331556\eta^4 - 1668.382017\eta^3 - 725.9372249\eta^2 - 218.6826746\eta + 1}{-682.0100712\eta^5 + 800.4008085\eta^4 - 79.09393360\eta^3 + 154.7934734\eta^2 - 222.6826746\eta + 1} \quad (16)$$

P[5/4] and P[5/5] have no positive real poles. The positive real poles of P[4/5] are 0.8737, 1.0237, and 1.8323; those of P[5/6] are 0.0930, 0.8521, 1.1325, and 1.5293. With the coefficients in Eq. (16), P[6/5] has a positive real pole at $\eta \approx 0.00450477037$. A numerator zero lies approximately $9.3 \times 10^{-15}$ away, making this an almost canceled pole–zero pair. We do not interpret it as a physical divergence.

Pieprzyk *et al.*[37] updated the Kolafa-Labik-Malijevský equation (KLM)[25]

$$Z_{\mathrm{KLM}} = 1 + 4x + 6x^2 + 2.3647684x^3 - 0.8698551x^4 + 1.1062803x^5 - 1.095049x^6 + 0.637614x^7 - 0.2279397x^{10} + 0.1098948x^{14} - 0.00906797x^{22} \quad (17)$$

to be a new form as[37]

$$Z_{\mathrm{mKLM}} = 1 + 4x + 6x^2 + 2.3647684x^3 - 0.8698551x^4 +$$

$$1.1062803x^5 - 1.1014221x^6 + 0.66605866x^7 - 0.03633431x^8 - 0.20965164x^{10} + 0.10555569x^{14} - 0.00872380x^{22} \quad (18)$$

by using their Molecular Dynamics simulation data. Compared with Eq. (17), an extra term of $x^8$ is added into Eq. (18). Here $x = \eta/(1-\eta)$. It is found[37] that Eq. (18) well stands up to their ending density $\rho = 1.02$ for the metastable region. $\eta_p$ of both KLM equation and mKLM equation is 1.

Based on a summation of the infinite sequence of virial coefficients, Hu and Yu[51] proposed two HS EOSs, namely HY1 and HY2, as follows:

$$Z_{HY1} = \sum_{n=1}^{m} B_n \eta^{n-1} + \frac{B_{m+1}\eta^m}{1-c\eta} \quad (19)$$

$$Z_{HY2} = 1 + \frac{B_2\eta}{1-c\eta} + \sum_{n=3}^{m} (B_n - B_2 c^{n-2})\eta^{n-1} \quad (20)$$

In the present comparison, HY1 denotes Eq. (19) and HY2 denotes Eq. (20). Both use m = 18 and $c = 1/\eta_c = \sqrt{18}/\pi$, together with the Hu–Yu [3/3] virial-coefficient approximation [Eq. (3) and Table 4 of Ref. 51]. Thus, both have the pole $\eta_c = \pi/\sqrt{18} \approx 0.74048049$. These fixed choices specify the two parameterizations evaluated in Tables 2 and 3.

In 2022, Davidchack and Laird successfully measured the excess chemical potential with high precision for the full range of HS particle diameter values from zero to infinity in HS fluids[11]. They also proposed an empirical HS EOS by the weighted least-squares fit to the combined simulation data from Kolafa *et al.*[25] and Pieprzyk *et al.*[37] and their own simulation data[11], namely DL, as follows[11]

$$Z_{DL} = 1 + 4x + 6x^2 + 2.36476838x^3 - 0.867384x^4 + 1.0598991x^5 - 0.8542136x^6 + 0.9972434x^8 - 0.8407042x^9 + 0.1507499x^{12} + 0.0532314x^{20} - 0.0469318x^{21} \quad (21)$$

In 2025, Skvara and Nezbeda proposed an accurate EOS for hard polyhedron fluids, namely SN, as follows[52]

$$Z_{SN} = \frac{1 + a_1\eta + a_2\eta^2 + a_3\eta^3 + \sum_{i=0}^{3} d_i \eta^{i+4}}{(1-\eta)^3} \quad (22)$$

Excellent agreement between the numerical values from Eq. (22) and the simulation data of the compressibility factor versus the packing fraction is achieved for all the polyhedron models considered. For the HS case, $a_1 = B_2 - 3 = 1, a_2 = B_3 - 3a_1 - 6 = 1, a_3 = -0.635232, d_0 = -4.347, d_1 = 27.997, d_2 = -71.984, d_3 = 61.086$.

The SN benchmark uses Eq. (13) and the Sphere row of Table I in Ref. 52. Its coefficients $a_1 \sim a_3$ enforce $B_2 \sim B_4$, whereas all four $d_0 \sim d_3$ were fitted to the simulation data of Irrgang et al.[36,52] This is distinct from fixing coefficients through $B_7$ and fitting only $d_3$. The present EOS in the following section instead prescribes $B_2 \sim B_{11}$, uses $B_{12} \sim B_{16}$ in its intermediate regression, and fits its remaining parameters to the dense Pieprzyk subset. Their different inputs preclude a controlled test of functional forms under identical constraints; the higher-order differences do not establish an intrinsic deficiency of the SN framework. Extending the SN coefficient-matching conditions would define a separate parameterization, which is not evaluated here.

These selected benchmarks illustrate different modeling priorities. The KLM, mKLM, and DL equations closely reproduce simulation pressures, with singularities at $\eta = 1$. Several virial-based approximants emphasize coefficient matching, while the Bonneville equation includes a pole near a representative disordered packing limit. Their differing inputs and analytic forms preclude interpreting every comparison as a test under identical constraints. Section 4 considers a specific combination of prescribed low-order inputs, an intermediate continuation, and an independently chosen representative pole. The intended benefit is this combination, rather than uniformly smaller pressure errors.

## 4. New hard sphere EOS

We retain the closed-virial framework of Tian et al.[53] The term $Z_T$ reproduces the prescribed low-order inputs, $Z_L$ supplies a fitted linear continuation over a finite range of orders, and $Z_I$ supplies a geometric tail with a prescribed pole. This is a semiempirical closure. The roles of its terms are explicit, but the selected continuation is not uniquely implied by hard-sphere statistical mechanics. The compressibility factor is

$$Z = Z_T + Z_L + Z_I \tag{23}$$

with

$$Z_T = 1 + B_2\eta + B_3\eta^2 + \cdots + B_n\eta^{n-1} \tag{24}$$

$$Z_L = \sum_{i=n}^{m} (c_1 + c_2 i)\,\eta^i \tag{25}$$

$$Z_I = \sum_{i=m+1}^{\infty} c_0\,\eta^i/\eta_p^i \tag{26}$$

The amplitude $c_0$ satisfies $c_0 = B_{m+2}\eta_p^{m+1}$ within this model, while $c_1$ and $c_2$ specify the linear coefficient sequence. This identity does not provide an independently known value of $B_{m+2}$. The earlier construction used $\eta_p = 0.7405$.[53]

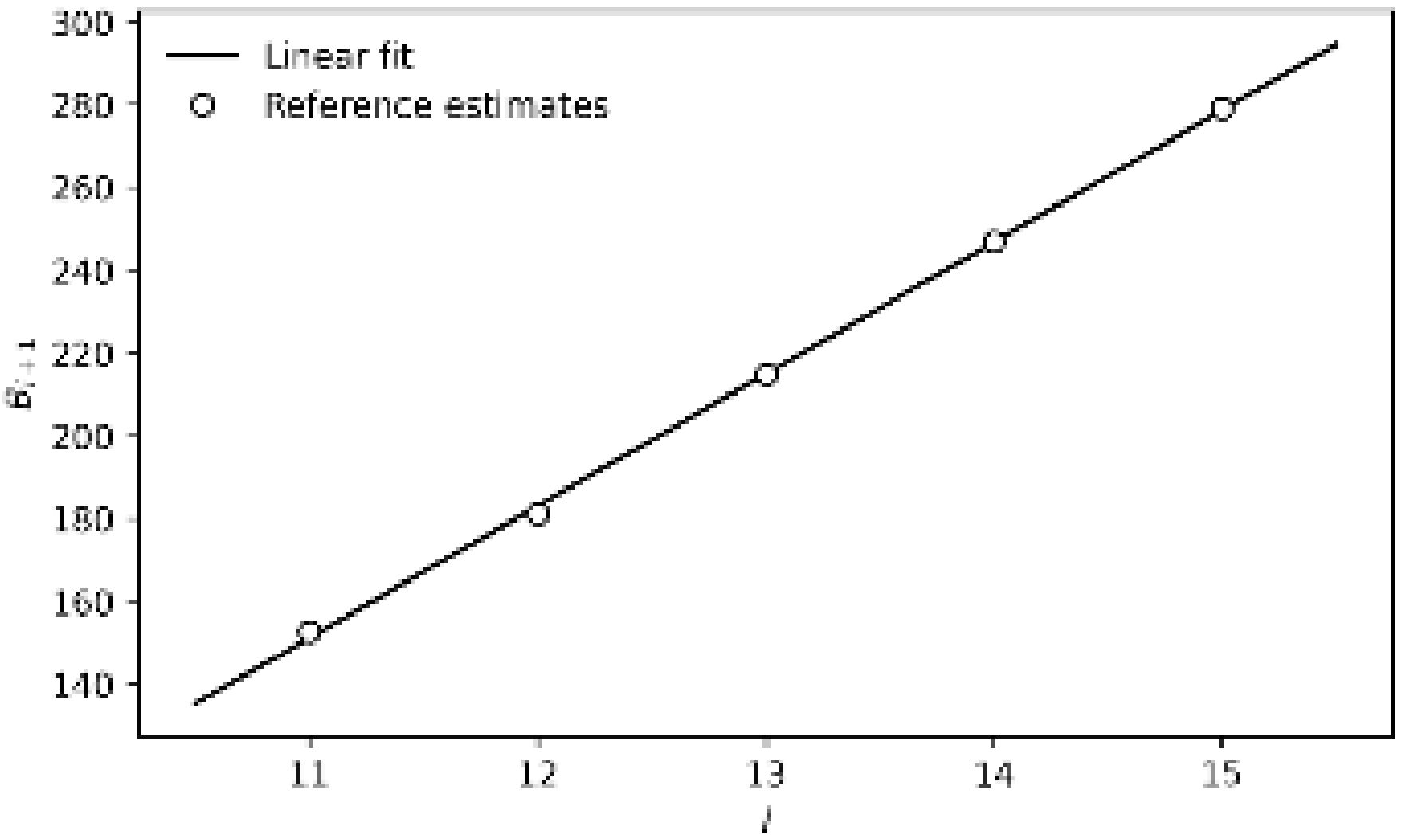

Figure 1. Linear fitting for Eq. (25) by using the predicted values of $B_{12}$-$B_{16}$ by Clisby and McCoy[22].

The closed form of Eq. (23) reads

$$Z = Z_T + \frac{(c_1 + c_2 n)\eta^n + (c_2 - c_1 - c_2 n)\eta^{n+1}}{(1-\eta)^2} + \frac{(-c_2 - c_1 - c_2 m)\eta^{m+1} + (c_1 + c_2 m)\eta^{m+2}}{(1-\eta)^2} + \frac{c_0 \eta^{m+1}}{\eta_p^{m+1}(1-\eta/\eta_p)}. \tag{27}$$

Equation (24) preserves the chosen inputs. Equation (25) adopts a finite linear coefficient continuation, motivated by the intermediate-order trend discussed in Ref. 54 and illustrated in Fig. 1. Equation (26) imposes a geometric tail consistent with the selected limiting ratio. That ratio alone does not determine the onset order or justify an exact geometric law from a finite order onward. Both are features of the closure, and the resulting unknown high-order coefficients are model quantities.

The earlier choice $\eta_p = 0.7405$ corresponds to crystalline close packing.[53] It is not the endpoint we seek to model for a disordered continuation. The equilibrium fluid freezes near $\eta_f \simeq 0.492$, and coexistence extends to approximately $\eta_m \simeq 0.543$; the stable higher-density branch is crystalline.[29,55] With crystallization suppressed, disordered states can instead approach jammed configurations. We prescribe $\eta_p = 0.637$ as a representative endpoint motivated by the jamming literature.[56–59] We do not derive this value from the pressure fit or identify it with a universal convergence radius of the equilibrium virial series. The MRJ concept concerns nonequilibrium jammed states, and the interpretation of a numerical endpoint depends on the disorder and jamming criteria. Fixing this representative pole is a modeling choice for the present hard-sphere continuation, not a universal prescription for other hard-body systems. Allowing a pole to vary in a different hard-body model addresses a different closure and fitting problem; transferability requires separate validation.

We set $n = 11$ and preserve the analytical values of $B_2$–$B_4$ and the specified numerical inputs of $B_5$–$B_{11}$.[8,9] The less certain higher-order information enters through a regression of the adopted $B_{12}$–$B_{16}$ estimates,[22] rather than by imposing $B_{12}$ as an exact polynomial coefficient. The fitted relation is

$$B_{i+1} = c_1 + c_2 i \tag{28}$$

with $c_1$=-199.4475 and $c_2$ =31.8767. The linear fitting is shown in **F**igure 1 here.

The five reference estimates do not establish how far the linear continuation remains appropriate. We determine the cutoff order $m$ and tail amplitude $c_0$ from pressure data. For reporting purposes, we retain the density groups used in the earlier analysis: a low-density group with $0.050 \le \rho^* \le 0.938$, a high-density group with $0.940 \le \rho^* \le 1.020$, and their union. These are numerical dataset partitions, not exact coexistence boundaries. The high-density group includes states on the metastable continuation. Relative deviations are defined by

$$\mathrm{PD}(i) = 100(Z_{\mathrm{EOS}}(i) - Z_{sim}(i))/Z_{sim}(i) \tag{29}$$

$$\mathrm{AAD} = \frac{1}{M}\sum_{i=1}^{M} |\mathrm{PD}(i)| \tag{30}$$

with PD the percentage deviation, $Z_{\mathrm{EOS}}$ the compressibility factor from EOS, $Z_{sim}$ the compressibility factor from the MD simulations and $M$ being the number of data points.

We define three AADs as

AAD1 denotes the mean absolute percentage deviation over the 59 low-density points.

AAD2 denotes the mean absolute percentage deviation over the 41 high-density points used for fitting.

AAD3 denotes the mean absolute percentage deviation over all 100 points.

The selected parameters are $m = 39$ and $c_0 = 15.2021$. Direct evaluation gives AAD1 = 0.021342%, AAD2 = 0.023826%, and AAD3 = 0.022361%. Thus, the explicit EOS is

$$\begin{aligned} Z = {} & 1 + 4\eta + 10\eta^2 + \frac{2707\pi + 438\sqrt{2} - 4131\arccos(1/3)}{70\pi}\eta^3 \\ & +28.224380\eta^4 + 39.81524\eta^5 + 53.3421\eta^6 + 68.529\eta^7 \\ & +85.825\eta^8 + 105.68\eta^9 + 127\eta^{10} \\ & + \frac{151.1962\eta^{11} - 119.3195\eta^{12}}{(1-\eta)^2} \\ & + \frac{-1075.6205\eta^{40} + 1043.7438\eta^{41}}{(1-\eta)^2} \end{aligned}$$

$$+\frac{15.2021\eta^{40}}{0.637^{40}(1-\eta/0.637)}.$$

(31)

The fit can be reproduced without a nonlinear amplitude search. For fixed $m$, write $Z_i = Z_{\mathrm{bg},i} + c_0 f_i$, where $Z_{\mathrm{bg}} = Z_T + Z_L$ and $f_i = \left(\eta_i/\eta_p\right)^{m+1}/\left(1-\eta_i/\eta_p\right)$. Defining $u_i = f_i/Z_{\mathrm{sim},i}$ and $v_i = \left(Z_{\mathrm{sim},i} - Z_{\mathrm{bg},i}\right)/f_i$, minimizing AAD2 is equivalent to minimizing $\sum_i u_i\,|c_0 - v_i|$. The amplitude is therefore a weighted median of $v_i$ with positive weights $u_i$, subject to $c_0 \geq 0$. A reproducibility check over integer $16 \leq m \leq 80$ selects $m = 39$, with $c_0 = 15.2020622$, consistent with the rounded value retained in Eq. (31). Table 4 gives the neighboring-order sensitivity. The interval searched is stated explicitly and does not establish optimality over every possible continuation.

The selected $m = 39$ minimizes AAD2 over the stated search interval. Table 4 shows that nearby larger m values can give slightly smaller AAD3; AAD3 was not the selection objective. This search compares parameters within the assumed closure and does not derive its functional form.

The regression range and the fitted cutoff play different roles. The available order-12–16 estimates constrain the line, whereas pressure data determine how long that line is retained in the approximation.

## 5. Results and analysis

The comparisons below assess three construction objectives. The pole and coefficient matching are properties of the approximation, while pressure agreement is assessed over specified datasets.

(1) A representative positive-density pole for the metastable disordered continuation.

(2) Preservation of the chosen low-order inputs and consistency with the adopted higher-order estimates.

(3) Small deviations from simulation pressures over the tested density intervals.

### 5.1 High-density divergence

Equation (31) has the prescribed pole at $\eta_p = 0.637$. The finite polynomial contributions remain regular there, while $Z_I$ diverges positively. The linear finite sum in Eq. (25) does not produce a genuine pole at $\eta = 1$; its denominator in Eq. (27) is removable when the numerator coefficients are retained consistently. The chosen disordered endpoint differs from the $\eta = 1$ singularities of KLM, mKLM, DL, and SN and from the crystalline endpoint used in HY1 and HY2. Bonneville includes a nearby pole at 0.6374.[47] These analytic differences are useful for defining the intended continuation, but a prescribed pole does not validate the pressure arbitrarily close to it. The largest packing fraction tested here is 0.53407075, well below 0.637. The exact equilibrium virial singularity and deep metastable behavior are not established by the present fit.

### 5.2 Virial-coefficient consistency

Equation (31) reproduces the specified $B_2$–$B_{11}$ inputs by construction. The estimates of $B_{12}$–$B_{16}$ enter the regression for $Z_L$, so their subsequent agreement is a consistency check rather than an independent prediction test. Table 2 compares the resulting coefficients with the adopted reference estimates. For Eq. (31), the absolute deviations in $B_{13}$–$B_{16}$ are approximately 1.04%, 0.09%, 0.05%, and 0.17%. The $B_{13}$ deviation exceeds the 0.93% relative uncertainty quoted in the table, so agreement within every listed uncertainty interval is not claimed.

The negative $B_{15}$ and $B_{16}$ of the EA12 expression disagree with the adopted positive reference estimates, but their signs alone do not demonstrate that the approximant is unphysical. Likewise, the finite-order decrease in EA11 coefficients is not a contradiction of an asymptotic ratio limit.

Direct expansion of the displayed DL and SN equations gives the corresponding entries in Table 2. For SN, $B_{13} = 279.59324$, $B_{14} = 371.838688$, $B_{15} = 479.200904$, and $B_{16} = 601.679888$. This calculation describes the stated parameterization; it does not turn the unequal-input comparison into a controlled test of functional forms. The same distinction applies when an EOS is assessed against high-order estimates that entered its construction.

### 5.3 Agreement with simulation data

Table 3 reports deviations from the Pieprzyk dataset using the same density groups. The directly recomputed values for mKLM are AAD1 = 0.000125%, AAD2 = 0.000198%, and AAD3 = 0.000155%; DL gives 0.000161%, 0.000216%, and 0.000183%, respectively. Evaluation of the displayed KLM equation, Eq. (17), gives 0.000156%, 0.000470%, and 0.000285%, respectively. Equation (31) gives 0.021342%, 0.023826%, and 0.022361%. Thus, all three empirical benchmarks are more accurate than Eq. (31) on the high-density subset and on the complete dataset. Since the principal dataset contributes to the construction of mKLM, DL, and the present EOS, these comparisons include calibration performance and cannot uniformly be described as independent predictions.

For CSK, AAD1, AAD2, and AAD3 are 0.020657%, 0.309389%, and 0.139037%, respectively. Relative to CS, CSK improves the lower-density and overall averages but has a larger deviation on the dense subset. Thus, the Kolafa modification does not improve every density partition.

With all parameters unchanged, we evaluate Eq. (31) on the 50 Davidchack–Laird points at $\eta = 0.01, 0.02, \ldots, 0.50$.[11] The AAD4 is 0.005874%, with a maximum absolute relative deviation of 0.049299%. Table 5 gives the corresponding directly evaluated values for DL, mKLM, and KLM: 0.00002375%, 0.00011647%, and 0.00015137%, respectively. The present equation therefore does not outperform these empirical benchmarks in pressure accuracy. Because the Davidchack–Laird data contributed to the DL fit, its residuals are not an independent-validation result. For Eq. (31), this is a fixed-parameter external assessment. It tests another dataset over overlapping densities and does not validate an extrapolation toward the imposed pole.

On the same Davidchack–Laird dataset, CSK gives AAD4 = 0.00888915%, an RMS relative deviation of 0.02259405%, and a maximum absolute relative deviation of 0.12794977% at $\eta = 0.50$ (Table 5). Its AAD4 is larger than that of Eq. (31).

## 6. Conclusions

We have compared selected hard-sphere equations and revised an existing closed-virial construction to include a representative disordered endpoint at $\eta_p = 0.637$. The equation preserves its specified low-order inputs and reproduces the higher-order estimates used in the intermediate regression with small deviations. These are construction properties and should be distinguished from prediction of unknown virial coefficients.

The overall pressure AAD of Eq. (31) is 0.022361% on the 100-point Pieprzyk dataset. Evaluation on 50 Davidchack–Laird points without refitting gives AAD4 = 0.005874% and a maximum deviation of 0.049299%. The mKLM, DL, and KLM equations have lower pressure errors over these tested datasets. Consequently, the contribution of Eq. (31) is the explicit combination of prescribed virial information, a finite-density disordered continuation, and a compact closed form, rather than an improvement in every numerical metric.

The high-order tail is phenomenological. Its geometric coefficients implement a selected asymptotic behavior, but the finite onset order is not derived from exact cluster integrals. Neither negative coefficients nor finite-order decreases in other approximants are rejected solely on those grounds. Comparisons with SN and other equations must also account for their different input information. The fitted parameter sensitivity provides a reproducibility check within the chosen model family, not proof that the family is unique.

Further progress requires better constrained high-order virial information and pressure data closer to the disordered packing limit. Such information could test the continuation and the assumed endpoint independently. The present EOS is supported over the examined density ranges, while its deep metastable extrapolation and extension to other hard-body systems remain open questions.

## Acknowledgments

The National Natural Science Foundation of China under Grant No. 11274200, the Natural Science Foundation of Shandong Province under Grant No. ZR2026MS0087 have supported this work. Tian would like to thank Dr. A. Mulero and Dr. I. Cachadina (Department of Applied Physics, University of Extremadura at Badajoz, Spain) for the helpful discussions and comments on this work, thank Dr. David A. Kofke (Department of Chemical and Biological Engineering, University at Buffalo, The State University of New York, USA) for his help on Table 1 of this paper, thank Dr. Alessio Zaccone (Department of Physics, University of Milan, Italy) and Dr. Ruslan Davidchack (School of Computing and Mathematical Sciences, University of Leicester, UK) for their helpful discussions and comments on this work. Cui shows her special thanks to her parents Mei and Bin for their deep love and strong support. The authors sincerely appreciate the anonymous reviewers for devoting their precious time to reviewing our work and their wonderful comments to update our work.

**Table 1**. Exact values, numerical estimates, and extrapolated estimates of the dimensionless virial coefficients $B_2$–$B_{16}$. Italic and bold entries identify the source selections described in Refs. 24 and 8, respectively. The extrapolated $B_{12}$–$B_{16}$ values are construction inputs, not exact cluster-integral results.

| **Virial coefficients** | **Exact/Numerical/predicted values** | **Standard errors** | **References** |
|---|---|---|---|
| Exact values | | | |
| $B_2$ | 4 | 0 | 1, 14-16 |
| $B_3$ | 10 | 0 | 1, 14-16 |
| $B_4$ | 18.364768… | 0 | 1, 14-16 |
| Numerical values | | | |
| $B_5$ | 28.224367 | 0.000017 | 24 |
| | 28.22441 | 0.00003 | 26 |
| | 28.22445 | 0.00010 | 21 |
| | 28.22445 | 0.00010 | 25 |
| | 28.2245 | 0.0003 | 22 |
| | *28.224377* | *0.000015* | 24 |
| | **28.224380** | **0.000015** | 8 |
| $B_6$ | 39.81524 | 0.00010 | 24 |
| | 39.8150 | 0.0002 | 26 |
| | 39.81547 | 0.00038 | 25 |
| | 39.81550 | 0.00036 | 21 |
| | 39.8151 | 0.0009 | 22 |
| | *39.81523* | *0.00010* | 24 |
| | **39.81524** | **0.00009** | 8 |
| $B_7$ | 53.3418 | 0.0005 | 24 |
| | 53.3435 | 0.0011 | 26 |
| | 53.3413 | 0.0016 | 21 |
| | 53.344 | 0.004 | 22 |
| | *53.3421* | *0.0005* | 24 |
| | **53.3421** | **0.0005** | 8 |
| $B_8$ | 68.526 | 0.003 | 24 |
| | 68.534 | 0.009 | 26 |
| | 68.540 | 0.010 | 21 |
| | 68.538 | 0.018 | 22 |
| | *68.526* | *0.003* | 24 |
| | **68.529** | **0.003** | 8 |

| **Virial coefficients** | **Exact/Numerical/predicted values** | **Standard errors** | **References** |
|---|---|---|---|
| $B_9$ | 85.83 | 0.02 | 24 |
| | 85.81 | 0.07 | 26 |
| | 85.80 | 0.08 | 21 |
| | 85.81 | 0.09 | 22 |
| | *85.83* | *0.02* | 24 |
| | **85.825** | **0.019** | 8 |
| $B_{10}$ | 105.64 | 0.10 | 24 |
| | 106.1 | 0.4 | 26 |
| | 106.2 | 1.0 | 23 |
| | 105.8 | 0.4 | 22 |
| | *105.68* | *0.10* | 24 |
| | **105.68** | **0.09** | 8 |
| $B_{11}$ | 126.4 | 0.6 | 24 |
| | 128 | 4 | 26 |
| | 128 | 5 | 23 |
| | *126.5* | *0.6* | 24 |
| | **127** | **3** | 8 |
| $B_{12}$ | 170 | 40 | 26 |
| | 111 | 30 | 23 |
| | *131* | *23* | 24 |
| | **133** | **23** | 8 |
| | 152.67 | **--** | 22 |
| Predicted values | | | |
| $B_{12}$ | 152.67 | -- | 22 |
| $B_{13}$ | 181.19 | -- | 22 |
| $B_{14}$ | 214.75 | -- | 22 |
| $B_{15}$ | 246.96 | -- | 22 |
| $B_{16}$ | 279.17 | -- | 22 |

**Table 2**. Comparison of $B_{13}$–$B_{16}$ with the reference estimates of Clisby and McCoy.[22] Parentheses in the reference row give the quoted relative uncertainties; parentheses in EOS rows give signed relative deviations where listed. Agreement with values used in fitting is not independent prediction. CM1 and CM2 denote Eqs. (4) and (5), respectively, of Ref. 14, which give the Clisby–McCoy Padé forms.[22] SH denotes the branch-point approximant, Eq. (2) with the $k = 3$ parameters in Table I of Ref. 50. YK denotes Eq. (4) of Ref. 35, with coefficient 40/3 multiplying $\eta^4$ in its numerator. These definitions also apply to Table 3.

| **EOS** | $B_{13}$ | $B_{14}$ | $B_{15}$ | $B_{16}$ | **Positive real poles** |
|---|---|---|---|---|---|
| Ref. ([22]) | 181.19 (0.93%) | 214.75 (3.1%) | 246.96 (1.1%) | 279.17 (3.9%) | -- |
| $Z_T$ | -- | -- | -- | -- | -- |
| $Z_T+Z_L$ | 183.0729 | 214.9496 | 246.8263 | 278.7030 | -- |
| Eq. (31) | 183.0729 (1.04%) | 214.9496 (0.09%) | 246.8263 (-0.05%) | 278.7030 (-0.17%) | 0.637 |
| Original $Z$ (-5,2) | 180.82 (-0.20%) | 212.56 (-1.02%) | 248.21 (0.51%) | 288.19 (3.23%) | 0.9262 |
| Updated $Z$ (-5,2) | 181.54 (0.19%) | 213.67 (-0.50%) | 249.84 (1.17%) | 290.47 (4.05%) | 0.9247 |
| $Z_{EA11}$ | 147.32 (-18.69%) | 143.63 (-33.12%) | 134.79 (-45.42%) | 120.18 (-56.95%) | -- |
| $Z_{EA12}$ | 101.94 (-43.74%) | 30.16 (-85.96%) | -73.60 (-129.80%) | -200.09 (-171.67%) | -- |
| $Z_{UC}$ | 220.04 (21.44%) | 268.77 (25.15%) | 326.30 (32.13%) | 394.06 (41.15%) | 0.8881 |
| $Z_H$ | 174.90 (-3.47%) | 201.99 (-5.94%) | 231.02 (-6.45%) | 262.01 (-6.15%) | 1 |
| $Z_B$ | 472.83 (160.96%) | 711.67 (231.39%) | 1083.10 (338.57%) | 1662.54 (495.53%) | 1, 0.6374 |
| CS | 180 (-0.66%) | 208 (-3.14%) | 238 (-3.63%) | 270 (-3.28%) | 1 |
| CSK | 168.3333 (-7.10%) | 193.3333 (-9.97%) | 220.0000 (-10.92%) | 248.3333 (-11.05%) | 1 |
| $Z_{KLM}$ | 197.74 (9.13%) | 260.05 (21.09%) | 330.61 (33.87%) | 361.62 (29.53%) | 1 |
| $Z_{mKLM}$ | 196.95 (8.70%) | 256.53 (19.46%) | 321.95 (30.37%) | 346.14 (23.99%) | 1 |
| YK | 284.12 | 378.83 | 505.11 | 673.48 | 0.75 |
| P[4/5] | 182.68 (0.82%) | 215.74 (0.46%) | 253.42 (2.62%) | 296.53 (6.22%) | 0.8737, 1.0237, 1.8323 |

| **EOS** | **$B_{13}$** | **$B_{14}$** | **$B_{15}$** | **$B_{16}$** | **Positive real poles** |
|---|---|---|---|---|---|
| P[5/4] | 177.11 (-2.25%) | 202.24 (-5.83%) | 229.91 (-6.90%) | 266.85 (-4.41%) | -- |
| P[5/5] | 177.06 (-2.28%) | 201.63 (-6.11%) | 229.23 (-7.18%) | 266.89 (-4.40%) | -- |
| P[5/6] | 972.40 | 8680.93 | 91228.15 | 978433.34 | 0.0930, 0.8521, 1.1325, 1.5293 |
| P[6/5] | $3.0*10^{16}$ | $7.0*10^{18}$ | $1.5*10^{21}$ | $3.0*10^{23}$ | 0.0045[a] |
| CM1, P[4/5] | 184.22 (1.67%) | 218.69 (1.83%) | 258.44 (4.65%) | 304.58 (9.10%) | 0.8580, 1.0983, 1.5790 |
| CM2, P[5/4] | 177.40 (-2.09%) | 203.23 (-5.36%) | 229.40 (-7.11%) | 267.73 (-4.10%) | -- |
| SH | 171.28 | 197.26 | 224.98 | 254.25 | 1 |
| HY1 | 174.49 (-3.70%) | 201.59 (-6.13%) | 230.69 (-6.59%) | 261.77 (-6.23%) | 0.7405 |
| HY2 | 174.49 (-3.70%) | 201.59 (-6.13%) | 230.69 (-6.59%) | 261.77 (-6.23%) | 0.7405 |
| $Z_{DL}$ | 202.58880 (+11.81%) | 264.84031 (+23.32%) | 317.87692 (+28.72%) | 291.06712 (+4.26%) | 1 |
| $Z_{SN}$ | 279.59324 (+54.31%) | 371.83869 (+73.15%) | 479.20090 (+94.04%) | 601.67989 (+115.52%) | 1 |

[a] Calculated from the coefficients in Eq. (16); the pole is almost canceled by a nearby numerator zero. The P[6/5] high-order coefficients consequently depend strongly on the precision of those coefficients.

**Table 3**. Mean absolute percentage deviations from the 100-point Pieprzyk dataset.[37] AAD1 uses 59 points with $0.050 \le \rho^* \le 0.938$, AAD2 uses 41 points with $0.940 \le \rho^* \le 1.020$, and AAD3 uses all points. These are dataset partitions, not exact phase boundaries. Additional digits are retained for the most accurate benchmarks.

| EOSs | AAD1 (%) | AAD2(%) | AAD3(%) | Construction information |
|---|---|---|---|---|
| $Z_T$ | 0.4178 | 1.8977 | 1.0245 | Prescribed $B_2$–$B_{11}$ |
| $Z_T$+$Z_L$ | 0.0183 | 0.1430 | 0.0694 | $B_2$–$B_{11}$; regression to $B_{12}$–$B_{16}$ |
| $Z_T$+$Z_L$+$Z_I$, Eq. (31) | 0.021342 | 0.023826 | 0.022361 | $B_2$–$B_{11}$; $B_{12}$–$B_{16}$ regression; fixed pole; dense Pieprzyk fit |
| updated Z(-5,2) | 0.0245 | 0.1222 | 0.0646 | Updated virial inputs; Refs. 8, 9 |
| EA11 | 0.0781 | 0.5757 | 0.2821 | Virial matching through $B_{11}$ |
| EA12 | 0.2370 | 1.3283 | 0.6844 | Virial matching through $B_{12}$ |
| UC | 0.0851 | 0.1503 | 0.1118 | Published coefficients; Ref. 45 |
| H | 0.0040 | 0.2030 | 0.0856 | Published coefficients; Ref. 46 |
| B | 2.2375 | 8.4011 | 4.7646 | Published coefficients; Ref. 47 |
| Original Z(-5,2) | 0.0217 | 0.1287 | 0.0656 | Original virial inputs; Ref. 14 |
| CS | 0.1515 | 0.2332 | 0.1850 | Parameter-free reference form |
| CSK | 0.020657 | 0.309389 | 0.139037 | Parameter-free Kolafa modification |
| KLM | 0.000156 | 0.000470 | 0.000285 | Virial constraints and simulation fit; Ref. 25 |
| mKLM | 0.000125 | 0.000198 | 0.000155 | Updated fit including Pieprzyk data; Ref. 37 |
| YK | 0.1857 | 0.7914 | 0.4340 | Published formula; Ref. 35, Eq. (4) |
| P[4/5] | 0.0300 | 0.1111 | 0.0632 | Published virial-based Padé coefficients; Ref. 8 |
| P[5/4] | 0.0049 | 0.1977 | 0.0839 | Published virial-based Padé coefficients; Ref. 8 |
| P[5/5] | 0.0044 | 0.2012 | 0.0851 | Published virial-based Padé coefficients; Ref. 8 |
| P[5/6] | 0.0425 | 0.0961 | 0.0645 | Published virial-based Padé coefficients; Ref. 8 |
| P[6/5] | 0.0049 | 0.1976 | 0.0839 | Published virial-based Padé coefficients; Ref. 8 |
| CM1 | 0.0365 | 0.1011 | 0.0630 | Published Padé coefficients; Ref. 14, Eq. (4) |
| CM2 | 0.0060 | 0.1919 | 0.0822 | Published Padé coefficients; Ref. 14, Eq. (5) |
| SH | 0.0195 | 0.2790 | 0.1259 | Published k = 3 parameter set; Ref. 50 |
| HY1 | 0.0063 | 0.2133 | 0.0912 | Hu–Yu [3/3] coefficients; m = 18; c = $\sqrt{18/\pi}$ |
| HY2 | 0.0034 | 0.1693 | 0.0714 | Hu–Yu [3/3] coefficients; m = 18; c = $\sqrt{18}/\pi$ |
| $Z_{DL}$ | 0.000161 | 0.000216 | 0.000183 | Combined simulation fit including both test datasets |
| $Z_{SN}$ | 0.040020 | 0.087407 | 0.059449 | $B_2$–$B_4$; four fitted d parameters; Irrgang data |

**Table 4**. Sensitivity to the cutoff order at a fixed pole of 0.637. The amplitude is refitted to the same 41 high-density points at each m. AADs are percentages. The selected order is 39.

| *m* | Refitted $c_0$ | AAD2 (%) | AAD3 (%) |
|---|---|---|---|
| 37 | 10.5191901 | 0.024128 | 0.022849 |
| 38 | 12.6456846 | 0.023876 | 0.022555 |
| 39 | 15.2020622 | 0.023826 | 0.022361 |
| 40 | 18.5863442 | 0.023931 | 0.022273 |
| 41 | 22.5932812 | 0.024152 | 0.022235 |

**Table 5**. Fixed-parameter evaluation on 50 Davidchack–Laird pressure points at packing fractions 0.01 to 0.50. All errors are percentages. DL includes these data in its fit and is a data-informed reference. RMS is the square root of the mean squared percentage deviation, and Maximum is the largest absolute percentage deviation: $RMS=\sqrt{\frac{1}{M}\sum_{i=1}^{M}[\mathrm{PD(i)}]^2}$ and $\mathrm{Maximum}=\max_i|\mathrm{PD}(i)|$, with $M$ = 50 and PD defined in Eq. (29).

| EOS | AAD4 (%) | RMS deviation (%) | Maximum (%) |
|---|---|---|---|
| DL | 0.00002375 | 0.00002890 | 0.00006402 |
| mKLM | 0.00011647 | 0.00014718 | 0.00030342 |
| KLM | 0.00015137 | 0.00020775 | 0.00055476 |
| Eq. (31) | 0.00587362 | 0.01458661 | 0.04929943 |
| CSK | 0.00888915 | 0.02259405 | 0.12794977 |